%% file: main.tex
\documentclass[aps,prx,twocolumn,floatfix,10pt,amssymb,amsfont,amsmath,superscriptaddress,float,nofootinbib,longbibliography]{revtex4-2}

\usepackage{preamb}
\input{tikz}

\begin{document}

\title{Low-depth unitary quantum circuits for dualities \\ in one-dimensional quantum lattice models}

\author{\textsc{Laurens Lootens}}
\email{laurens.lootens@ugent.be}
\affiliation{Department of Applied Mathematics and Theoretical Physics, University of Cambridge,\\ Wilberforce Road, Cambridge, CB3 0WA, United Kingdom}
\affiliation{Department of Physics and Astronomy, Ghent University, Krijgslaan 281, 9000 Gent, Belgium}
\author{\textsc{Clement Delcamp}}
\email{delcamp@ihes.fr}
\affiliation{Department of Physics and Astronomy, Ghent University, Krijgslaan 281, 9000 Gent, Belgium}
\affiliation{Institut des Hautes \'Etudes Scientifiques, Bures-sur-Yvette, France}
\author{\textsc{Dominic Williamson}}
\affiliation{Centre for Engineered Quantum Systems, School of Physics, University of Sydney, Sydney, New South Wales 2006, Australia}
\author{\textsc{Frank Verstraete}}
\affiliation{Department of Applied Mathematics and Theoretical Physics, University of Cambridge,\\ Wilberforce Road, Cambridge, CB3 0WA, United Kingdom}
\affiliation{Department of Physics and Astronomy, Ghent University, Krijgslaan 281, 9000 Gent, Belgium}

\begin{abstract}

\noindent
A systematic approach to dualities in symmetric (1+1)d quantum lattice models has recently been proposed in terms of module categories over the symmetry fusion categories. By characterizing the non-trivial way in which dualities intertwine closed boundary conditions and charge sectors, these can be implemented by unitary matrix product operators. In this manuscript, we explain how to turn such duality operators into unitary linear depth quantum circuits via the introduction of ancillary degrees of freedom that keep track of the various sectors. The linear depth is consistent with the fact that these dualities change the phase of the states on which they act. When supplemented with measurements, we show that dualities with respect to symmetries encoded into nilpotent fusion categories can be realised in constant depth. The resulting circuits can for instance be used to efficiently prepare short- and long-range entangled states or map between different gapped boundaries of (2+1)d topological models.

\end{abstract}

\maketitle

\input{_Introduction}

\input{_Dualities}

\input{_Linear.tex}

\input{_Constant}

\input{_Discussion}

\medskip\noindent
\textbf{Acknowledgments:} \emph{We would like to thank Gerardo Ortiz and Bram Vancraeynest-De Cuiper for previous collaborations, as well as Ruben Verresen and Nathanan Tantivasadakarn for inspiring discussions. This work has received funding from the Research Foundation Flanders (FWO) through postdoctoral fellowship No. 1228522N awarded to CD and doctoral fellowship No. 1184722N awarded to LL, and grants no. G0E1820 and G0E1520N.
}

\bigskip

\input{main.bbl}
\end{document}

%% file: tikz.tex
\pgfdeclarelayer{back}
\pgfdeclarelayer{front}
\pgfsetlayers{back,main,front}

\tikzset{->-/.style={decoration={
			markings,
			mark=at position #1 with {\arrow{Triangle[width=3.5pt, length=3.5pt]}}},postaction={decorate}}}

\tikzset{mask point/.style={ transform shape, sloped, 
minimum width=20pt, minimum height=4pt, inner sep=0pt, ultra thin, font=\tiny}}

\tikzset{
    clip even odd rule/.code={\pgfseteorule},
    invclip/.style={clip,insert path=[clip even odd rule]{
    [reset cm](-\maxdimen,-\maxdimen)rectangle(\maxdimen,\maxdimen)
    }}} 
    
\newcommand\clipIntersection[1]{
    (#1.north west) -- (#1.north east) -- (#1.south east) -- (#1.south west) -- (#1.north west)
}

\tikzstyle{dot}=[dash pattern= on 0.6pt off 2.1pt]
\tikzstyle{obj1}=[line width = 0.7pt, line join=round, line cap=round, shorten >= 1pt, shorten <= 1pt]
\tikzstyle{obj3}=[line width = 0.7pt, line join=round, line cap=round, shorten >= 1pt, shorten <= 1pt, color=gray]
\tikzstyle{obj2}=[line width = 0.7pt, line join=round, line cap=round, shorten >= 1pt, shorten <= 1pt, color=BrickRed]
\tikzstyle{mod}=[color=BlueViolet!80,line width = 0.7pt, line join=round, line cap=round]
\tikzstyle{mor}=[draw=none, fill=gray!22]
\tikzstyle{wig}=[snake=zigzag, segment amplitude=0.7pt, segment length=3.5pt, rounded corners=0.6pt]
\tikzstyle{ontop}=[fill=white, rectangle, inner sep=1pt, text=black]

\newcommand\clipAroundNode[2]{
    \begin{pgfscope}
        \begin{scope}[overlay]
            \path[invclip] \clipIntersection{#1};#2
        \end{scope}
    \end{pgfscope}
}

\newcommand\clipAroundTwoNodes[3]{
    \begin{pgfscope}
        \begin{scope}[overlay]
            \path[invclip] \clipIntersection{#1} \clipIntersection{#2};#3
        \end{scope}
    \end{pgfscope}
}

\newcommand\clipAroundThreeNodes[4]{
    \begin{pgfscope}
        \begin{scope}[overlay]
            \path[invclip] \clipIntersection{#1} \clipIntersection{#2} \clipIntersection{#3};#4
        \end{scope}
    \end{pgfscope}
}

\newcommand{\middleCoo}[6]{
    \path (#1) --coordinate[pos=0.5](#3) (#2);
    \path (#1) --coordinate[pos={0.5-\sp}](#4) (#2);
    \path (#1) --coordinate[pos={0.5+\sp}](#5) (#2);
}

\newcommand{\barycentre}[8]{
    \middleCoo{#1}{#2}{A}{}{}{}{#8};
    \middleCoo{#2}{#3}{B}{}{}{}{#8};
    \middleCoo{#1}{#3}{C}{}{}{}{#8};
    \path[name path=PA#3] (A) -- (#3);
    \path[name path=PB#1] (B) -- (#1);
    \path[name path=PC#2] (C) -- (#2);
    \path [name intersections={of=PA#3 and PB#1,by={#4}}];
    \path (#1) --coordinate[pos=0.84](#5) (#4);
    \path (#2) --coordinate[pos=0.84](#6) (#4);
    \path (#3) --coordinate[pos=0.84](#7) (#4);
}

\newcommand{\barycentreSq}[9]{
    \middleCoo{#1}{#2}{A1}{A2}{A3}{};
    \middleCoo{#3}{#4}{B1}{B2}{B3}{};
    \middleCoo{#1}{#3}{C1}{C2}{C3}{};
    \middleCoo{#2}{#4}{D1}{D2}{D3}{};
    \middleCoo{A1}{B1}{#5}{}{};
    \path[name path=PA2B2] (A2) -- (B2);
    \path[name path=PA3B3] (A3) -- (B3);
    \path[name path=PC2D2] (C2) -- (D2);
    \path[name path=PC3D3] (C3) -- (D3);       
    \path [name intersections={of=PA2B2 and PC2D2,by={#6}}];
    \path [name intersections={of=PA2B2 and PC3D3,by={#8}}];
    \path [name intersections={of=PA3B3 and PC2D2,by={#7}}];
    \path [name intersections={of=PA3B3 and PC3D3,by={#9}}];
}

\newcommand{\blob}[6]{
    \ifthenelse{#6=1}{
        \draw[fill=white,rounded corners=1pt] ($(#1,#2)+(-1.2*#3/2,-#4/2)$) rectangle ($(#1,#2)+(1.2*#3/2,#4/2)$);
    }{}
    \ifthenelse{#6=2}{
        \draw[BrickRed, fill=BrickRed!10,rounded corners=1pt] ($(#1,#2)+(-1.2*#3/2,-#4/2)$) rectangle ($(#1,#2)+(1.2*#3/2,#4/2)$);
    }{}
    \node at (#1,#2) {${\sss #5}$};
}

\newcommand{\splitSpace}[5]{
	\begin{tikzpicture}[scale=1.15,baseline=-13pt]
    	\def\l{3.5};
    	\def\sp{0.16};
    	\ifthenelse{#5=1}{
        	\draw[obj1, ->-=0.5] (0,-0.7) -- (0,-0.1);
        	\node at (0,-.8) {${\sss #2}$};
        	\draw[mor] ({-\sp*\l/2},0) to[out=-90, in=-90, looseness=1.5] ({\sp*\l/2},0);
        	\draw[mod] ({-\sp*\l/2},-0.7) --node[text=black,pos=0.5, xshift=-7pt](){${\sss #1}$} ({-\sp*\l/2},0);
        	\draw[mod] ({\sp*\l/2},-0.7) --node[text=black,pos=0.5, xshift=7pt](){${\sss #3}$} ({\sp*\l/2},0);
        	\node[yshift=-3pt] at (0,0) {${\sss #4}$};
    	}{}
    	\ifthenelse{#5=2}{
        	\draw[obj1, ->-=0.7] (0,-0.6) -- (0,0);
        	\node at (0,.1) {${\sss #2}$};
        	\draw[mor] ({-\sp*\l/2},-0.7) to[out=90, in=90, looseness=1.5] ({\sp*\l/2},-0.7);
        	\draw[mod] ({-\sp*\l/2},-0.7) --node[text=black,pos=0.5, xshift=-7pt](){${\sss #1}$} ({-\sp*\l/2},0);
        	\draw[mod] ({\sp*\l/2},-0.7) --node[text=black,pos=0.5, xshift=7pt](){${\sss #3}$} ({\sp*\l/2},0);
        	\node[yshift=3pt] at (0,-0.7) {${\sss #4}$};
    	}{}
	\end{tikzpicture}
}

\newcommand{\MPO}[6]{
    \def\styA{#1}
    \def\styB{#2}
    \def\morI{#3}
    \def\morII{#4}
    \def\morIII{#5}
    \def\morIV{#6}
    \MPOcont
}

\newcommand{\MPOcont}[7]{
	\begin{tikzpicture}[scale=1.15,baseline={([yshift=-.5ex]current bounding box.center)}]
        \def\r{8pt};
        \def\l{3.5};
        \def\sp{0.16};
        \coordinate (a1) at (0,0);
        \coordinate (a2) at (\l/2,0);
        \coordinate (a3) at (0,\l/2);
        \coordinate (a4) at (\l/2,\l/2);
        
        \middleCoo{a1}{a2}{m12}{m12-1}{m12-2}{\sp};
        \middleCoo{a1}{a3}{m13}{m13-1}{m13-3}{\sp};
        \middleCoo{a2}{a4}{m24}{m24-2}{m24-4}{\sp};
        \middleCoo{a3}{a4}{m34}{m34-3}{m34-4}{\sp};
        \barycentreSq{a1}{a2}{a3}{a4}{b1234}{b1234-1}{b1234-2}{b1234-3}{b1234-4};
        
        \ifthenelse{#7=3 \OR #7=5}{
            \draw[obj1,->-=0.7,shorten <= 1pt, shorten >= 1pt] (m13) --node[pos=0.35,fill=white,rectangle,inner sep=0pt,text=black] {${\sss #6}$} node[mask point, pos=0.5](mk1){}  (m24);
        }{}
        \ifthenelse{#7 = 4}{
            \draw[mod,->-=0.7,shorten <= 1pt, shorten >= 1pt] (m13) --node[pos=0.35,fill=white,rectangle,inner sep=0pt,text=black] {${\sss #6}$} node[mask point, pos=0.5](mk1){}  (m24);
        }{}
        \ifthenelse{#7 = 6}{
            \draw[obj2,->-=0.7,shorten <= 1pt, shorten >= 1pt] (m13) --node[pos=0.35,fill=white,rectangle,inner sep=0pt,text=black] {${\sss #6}$} node[mask point, pos=0.5](mk1){}  (m24);
        }{}
        \ifthenelse{#7 = 0}{
            \draw[mod,->-=0.7,shorten <= 1pt, shorten >= 1pt] (m24) --node[pos=0.35,fill=white,rectangle,inner sep=0pt,text=black] {${\sss #6}$} node[mask point, pos=0.5](mk1){}  (m13);
        }{}
        \ifthenelse{#7 = 1}{
            \draw[obj2,->-=0.7] (m24) --node[pos=0.35,fill=white,rectangle,inner sep=0pt,text=black] {${\sss #6}$} node[mask point, pos=0.5](mk1){}  (m13);
        }{}
        \ifthenelse{#7 = 2}{
            \draw[obj1, dot] (m24) -- node[mask point, pos=0.5](mk1){}  (m13);
        }{}

        \clipAroundNode{mk1}{
            \ifthenelse{#7=5}{
                \draw[obj2,->-=0.75] (m12) --node[pos=0.3,fill=white,rectangle,inner sep=1pt,text=black] {${\sss #5}$} (m34);
            }{
                \draw[obj1,->-=0.75] (m12) --node[pos=0.3,fill=white,rectangle,inner sep=1pt] {${\sss #5}$} (m34);
            }
        }  

        \draw[mor] 
        (m12-1) to[out=90, in=90, looseness=1.5] (m12-2)
        (m13-1) to[out=0, in=0, looseness=1.5] (m13-3)
        (m24-2) to[out=-180, in=-180, looseness=1.5] (m24-4)
        (m34-3) to[out=-90, in=-90, looseness=1.5] (m34-4);
        
        \draw[obj1, rounded corners=\r, \styA] 
        (m12-1) -- (b1234-1) -- (m13-1)
        (m12-2) -- (b1234-2) -- (m24-2);
        \draw[obj1, rounded corners=\r, \styB] 
        (m13-3) -- (b1234-3) -- (m34-3)
        (m34-4) -- (b1234-4) -- (m24-4);
        
        \node[yshift=3pt] at (m12) {${\sss \morIV}$};
        \node[yshift=-3.5pt] at (m34) {${\sss \morIII}$};
        \node[xshift=3pt] at (m13) {${\sss \morI}$};   
        \node[xshift=-3pt] at (m24) {${\sss \morII}$};   
        \node[xshift=-8pt,yshift=8pt] at (b1234-3) {${\sss #1}$};
        \node[xshift=8pt,yshift=8pt] at (b1234-4) {${\sss #2}$};
        \node[xshift=-8pt,yshift=-8pt] at (b1234-1) {${\sss #4}$};
        \node[xshift=8pt,yshift=-8pt] at (b1234-2) {${\sss #3}$}; 
	\end{tikzpicture}
}

\newcommand{\intertwiner}[1]{
    \begin{tikzpicture}[scale=1.15,baseline={([yshift=-.5ex]current bounding box.center)}]
        \def\r{8pt};
        \def\l{3.5};
        \def\sp{0.16};
        
        \foreach \x [count=\xi] in {0,\l/2,\l}{
            \coordinate (a1) at (\x,0);
            \coordinate (a2) at (\l/2+\x,0);
            \coordinate (a3) at (\x,\l/2+0);
            \coordinate (a4) at (\l/2+\x,\l/2+0);
            \middleCoo{a1}{a2}{m12-\xi}{m12-1-\xi}{m12-2-\xi}{\sp};
            \middleCoo{a1}{a3}{m13-\xi}{m13-1-\xi}{m13-3-\xi}{\sp};
            \middleCoo{a2}{a4}{m24-\xi}{m24-2-\xi}{m24-4-\xi}{\sp};
            \middleCoo{a3}{a4}{m34-\xi}{m34-3-\xi}{m34-4-\xi}{\sp};
            \barycentreSq{a1}{a2}{a3}{a4}{b1234-\xi}{b1234-1-\xi}{b1234-2-\xi}{b1234-3-\xi}{b1234-4-\xi};
        }   
        
        \ifthenelse{#1=1}{                  
            \draw[obj2,->-=0.65] (m24-3) --node[pos=0.35,fill=white,rectangle,inner sep=0pt,text=black] {${\sss X}$} node[mask point, pos=0.2](mk1){} node[mask point, pos=0.5](mk2){} node[mask point, pos=0.8](mk3){} (m13-1);
        }{}
        \ifthenelse{#1=2}{                  
            \draw[obj1,dot] (m24-3) --node[mask point, pos=0.2](mk1){} node[mask point, pos=0.5](mk2){} node[mask point, pos=0.8](mk3){} (m13-1);
        }{}
        
        \clipAroundThreeNodes{mk1}{mk2}{mk3}{
            \draw[obj1,->-=0.75] (m12-1) --node[pos=0.3,fill=white,rectangle,inner sep=0pt] {${\sss Y_{\msf i-\frac{1}{2}}}$} (m34-1);
            \draw[obj1,->-=0.75] (m12-2) --node[pos=0.3,fill=white,rectangle,inner sep=1pt] {${\sss Y_{\msf i+\frac{1}{2}}}$} (m34-2);
            \draw[obj1,->-=0.75] (m12-3) --node[pos=0.3,fill=white,rectangle,inner sep=1pt] {${\sss Y_{\msf i+\frac{3}{2}}}$} (m34-3);
        }  
        
        \foreach \xi in {1,2,3}{
            \draw[mor] 
                (m12-1-\xi) to[out=90, in=90, looseness=1.5] (m12-2-\xi)
                (m34-3-\xi) to[out=-90, in=-90, looseness=1.5] (m34-4-\xi);
            \ifthenelse{#1=1}{
                \draw[mod,rounded corners=\r] 
                    (m12-1-\xi) -- (b1234-1-\xi) -- (m13-1-\xi)
                    (m12-2-\xi) -- (b1234-2-\xi) -- (m24-2-\xi);
                \draw[mod,rounded corners=\r]     
                    (m34-3-\xi) -- (b1234-3-\xi) -- (m13-3-\xi)
                    (m34-4-\xi) -- (b1234-4-\xi) -- (m24-4-\xi);
            }{}
            \ifthenelse{#1=2}{
                \draw[mod,color=black,rounded corners=\r] 
                    (m12-1-\xi) -- (b1234-1-\xi) -- (m13-1-\xi)
                    (m12-2-\xi) -- (b1234-2-\xi) -- (m24-2-\xi);
                \draw[mod,color=black,dot,rounded corners=\r]     
                    (m34-3-\xi) -- (b1234-3-\xi) -- (m13-3-\xi)
                    (m34-4-\xi) -- (b1234-4-\xi) -- (m24-4-\xi);
            }{}
        }
        
        \ifthenelse{#1=1}{
            \draw[mod, dash pattern= on 0.2pt off 2.3pt] 
                (m24-2-3) -- ($(m24-2-3)+(0.3,0)$)
                (m24-4-3) -- ($(m24-4-3)+(0.3,0)$)
                (m13-1-1) -- ($(m13-1-1)+(-0.3,0)$)
                (m13-3-1) -- ($(m13-3-1)+(-0.3,0)$);
            \draw[obj2, dash pattern= on 0.2pt off 2.3pt] 
                (m24-3) -- ($(m24-3)+(0.3,0)$)
                (m13-1) -- ($(m13-1)+(-0.3,0)$);
        }{}

        \node[above] at (m13-3-1) {${\sss N_{\msf i-1}}$};
        \node[above] at (m13-3-2) {${\sss N_{\msf i}}$};
        \node[above] at (m13-3-3) {${\sss N_{\msf i+1}}$};
        \node[above] at (m24-4-3) {${\sss N_{\msf i+2}}$};
        \node[below] at (m13-1-1) {${\sss M_{\msf i-1}}$};
        \node[below] at (m13-1-2) {${\sss M_{\msf i}}$};
        \node[below] at (m13-1-3) {${\sss M_{\msf i+1}}$};
        \node[below] at (m24-2-3) {${\sss M_{\msf i+2}}$};
    \end{tikzpicture}
}

\newcommand{\tube}[2]{
    \def\morI{#1}
    \def\morII{#2}
    \tubeCont
}

\newcommand{\tubeCont}[8]{
    \def\modI{#1}
    \def\modII{#2}
    \def\modIII{#3}
    \def\modIV{#4}
    \def\modV{#5}
    \def\modVI{#6}
    \def\modVII{#7}
    \def\modVIII{#8}
    \tubeContCont
}

\newcommand{\tubeContCont}[7]{
    \begin{tikzpicture}[scale=1.15,baseline={([yshift=-.5ex]current bounding box.center)}]
        \def\r{8pt};
        \def\l{3.5};
        \def\sp{0.16};
        
        \foreach \x [count=\xi] in {0,3*\l/8,5*\l/8,\l}{
            \coordinate (a1) at (\x,0);
            \coordinate (a2) at (\l/2+\x,0);
            \coordinate (a3) at (\x,\l/2+0);
            \coordinate (a4) at (\l/2+\x,\l/2+0);
            \middleCoo{a1}{a2}{m12-\xi}{m12-1-\xi}{m12-2-\xi}{\sp};
            \middleCoo{a1}{a3}{m13-\xi}{m13-1-\xi}{m13-3-\xi}{\sp};
            \middleCoo{a2}{a4}{m24-\xi}{m24-2-\xi}{m24-4-\xi}{\sp};
            \middleCoo{a3}{a4}{m34-\xi}{m34-3-\xi}{m34-4-\xi}{\sp};
            \barycentreSq{a1}{a2}{a3}{a4}{b1234-\xi}{b1234-1-\xi}{b1234-2-\xi}{b1234-3-\xi}{b1234-4-\xi};
        }
        
        \draw[obj2, ->-=0.8] (b1234-2) -- node[mask point, pos=0.6](mk1){} node[pos=0.43,fill=white,rectangle,inner sep=0pt,text=black] {$\sss #1$} (m13-1);
        \draw[obj2, ->-=.75] (b1234-3) -- node[pos=0.41,fill=white,rectangle,inner sep=-1pt,text=black] {$\sss #2$} (b1234-2);
        \draw[obj2, ->-=0.6] (m24-4) -- node[pos=0.23,fill=white,rectangle,inner sep=0pt,text=black] {$\sss #3$} node[mask point, pos=0.4](mk2){} (b1234-3);
                        
        \draw[obj2,->-=0.7] (b1234-2) --node[pos=0.38,fill=white,rectangle,inner sep=1pt,text=black] {${\sss #4}$} (m34-2);
        \draw[obj2,->-=0.77] (m12-3) --node[pos=0.45,fill=white,rectangle,inner sep=1pt,text=black] {${\sss #5}$} (b1234-3);
        
        \clipAroundTwoNodes{mk1}{mk2}{
            \draw[obj1,->-=0.75] (m12-1) --node[pos=0.3,fill=white,rectangle,inner sep=0pt] {${\sss #6}$} (m34-1);
            \draw[obj1,->-=0.75] (m12-4) --node[pos=0.3,fill=white,rectangle,inner sep=0pt] {${\sss #7}$} (m34-4);
        }
        
        \draw[mor] 
            (m34-3-1) to[out=-90, in=-90, looseness=1.5] (m34-4-1)
            (m34-3-2) to[out=-90, in=-90, looseness=1.5] (m34-4-2)
            (m34-3-4) to[out=-90, in=-90, looseness=1.5] (m34-4-4)
            (m12-1-1) to[out=90, in=90, looseness=1.5] (m12-2-1)
            (m12-1-3) to[out=90, in=90, looseness=1.5] (m12-2-3)
            (m12-1-4) to[out=90, in=90, looseness=1.5] (m12-2-4);
                
        \draw[mod, rounded corners=\r] 
            (m34-3-1) -- (b1234-3-1) -- (m13-3-1)
            (m34-4-1) -- (b1234-4-1) -- ($(m24-4-1)+(-\l/16,0)$)
            (m34-3-4) -- (b1234-3-4) -- ($(m13-3-4)+(0,0)$)
            (m34-4-4) -- (b1234-4-4) -- ($(m24-4-4)+(0,0)$)
            (m12-1-1) -- (b1234-1-1) -- (m13-1-1)
            (m12-2-1) -- (b1234-2-1) -- ($(m24-2-1)+(0,0)$)
            (m12-1-4) -- (b1234-1-4) -- ($(m13-1-4)+(\l/16,0)$)
            (m12-2-4) -- (b1234-2-4) -- ($(m24-2-4)+(0,0)$)
            (m34-3-2) -- (b1234-3-2) -- ($(m24-4-1)+(-\l/16,0)$)
            (m34-4-2) -- (b1234-4-2) -- (m13-3-4)
            (m12-2-3) -- (b1234-2-3) -- ($(m13-1-4)+(\l/16,0)$)
            (m12-1-3) -- (b1234-1-3) -- (m24-2-1);
    
        \draw[mod, dash pattern= on 0.2pt off 2.3pt] 
            (m24-2-4) -- ($(m24-2-4)+(0.3,0)$)
            (m24-4-4) -- ($(m24-4-4)+(0.3,0)$)
            (m13-1-1) -- ($(m13-1-1)+(-0.3,0)$)
            (m13-3-1) -- ($(m13-3-1)+(-0.3,0)$);
        \draw[obj2, dash pattern= on 0.2pt off 2.3pt] 
            (m24-4) -- ($(m24-4)+(0.3,0)$)
            (m13-1) -- ($(m13-1)+(-0.3,0)$);
                
        \node[mor, circle, inner sep=4pt] at (b1234-2) {};
        \node at (b1234-2) {${\sss \morI}$};
        \node[mor, circle, inner sep=4pt] at (b1234-3) {};
        \node at (b1234-3) {${\sss \morII}$};
        \node[above] at (m13-3-1) {${\sss \modI}$};
        \node[below] at (m13-1-1) {${\sss \modV}$};
        \node[above] at ($(m24-4-1)+(-\l/16,0)$) {${\sss \modII}$};
        \node[below] at ($(m13-1-4)+(\l/16,0)$) {${\sss \modVII}$};
        \node[above] at ($(m24-4-2)+(\l/16,0)$) {${\sss \modIII}$};
        \node[below] at ($(m13-1-3)+(-\l/16,0)$) {${\sss \modVI}$};
        \node[above] at (m24-4-4) {${\sss \modIV}$};
        \node[below] at (m24-2-4) {${\sss \modVIII}$};
    \end{tikzpicture}
}

\newcommand{\infChain}[3]{
    \def\morI{#1}
    \def\morII{#2}
    \def\morIII{#3}
    \infChainCont
}

\newcommand{\infChainCont}[7]{
    \begin{tikzpicture}[scale=1.15,baseline={([yshift=-1ex]current bounding box.center)}]
        \def\r{8pt};
        \def\l{3.5};
        \def\sp{0.16};
        \coordinate (a1) at (0,0);
        \coordinate (a2) at (\l/2,0);
        \coordinate (a3) at (0,\l/2);
        \coordinate (a4) at (\l/2,\l/2);
    
        \clip (-0.55,\l/4-0.2) rectangle (3*\l/2+0.55,\l/2+0.1);
        
        \middleCoo{a1}{a2}{m12}{m12-1}{m12-2}{\sp};
        \middleCoo{a1}{a3}{m13}{m13-1}{m13-3}{\sp};
        \middleCoo{a2}{a4}{m24}{m24-2}{m24-4}{\sp};
        \middleCoo{a3}{a4}{m34}{m34-3}{m34-4}{\sp};
        \barycentreSq{a1}{a2}{a3}{a4}{b1234}{b1234-1}{b1234-2}{b1234-3}{b1234-4};
        
        \draw[obj1, ->-=0.6, shorten <= 7pt] ($(b1234)+(0,0)$) -- ($(m34)+(0,0)$);
        \draw[obj1, ->-=0.6, shorten <= 7pt] ($(b1234)+(\l,0)$) -- ($(m34)+(\l,0)$);
        \draw[obj1, ->-=0.6, shorten <= 7pt] ($(b1234)+(\l/2,0)$) -- ($(m34)+(\l/2,0)$);
        \foreach \x in {0,\l/2,\l}{
            \draw[mor] ($(m34-3)+(\x,0)$) to[out=-90, in=-90, looseness=1.5] ($(m34-4)+(\x,0)$);
            \draw[mod, rounded corners=\r] 
                ($(m13-3)+(\x,0)$) -- ($(b1234-3)+(\x,0)$) -- ($(m34-3)+(\x,0)$)
                ($(m24-4)+(\x,0)$) -- ($(b1234-4)+(\x,0)$) -- ($(m34-4)+(\x,0)$);
        }
        \draw[mod, dash pattern= on 0.2pt off 2.3pt] 
            (m13-3) -- ($(m13-3)+(-0.3,0)$)
            ($(m24-4)+(\l,0)$) -- ($(m24-4)+(\l+0.3,0)$);
        \node[yshift=1pt] at (\l/4,\l/4) {${\sss #5}$};
        \node[yshift=1pt] at (\l/4+\l/2,\l/4) {${\sss #6}$};
        \node[yshift=1pt] at (\l/4+\l,\l/4) {${\sss #7}$};
        \node[above] at (m13-3) {${\sss #1}$};
        \node[above] at ($(m13-3)+(\l/2,0)$) {${\sss #2}$};
        \node[above] at ($(m13-3)+(\l,0)$) {${\sss #3}$};
        \node[above] at ($(m13-3)+(3*\l/2,0)$) {${\sss #4}$};
        \node[yshift=-3.5pt] at ($(m34)+(0,0)$) {${\sss \morI}$};
        \node[yshift=-3.5pt] at ($(m34)+(\l/2,0)$) {${\sss \morII}$};
        \node[yshift=-3.5pt] at ($(m34)+(\l,0)$) {${\sss \morIII}$};
    \end{tikzpicture}
}

\newcommand{\chain}[3]{
    \def\morI{#1}
    \def\morII{#2}
    \def\morIII{#3}
    \chainCont
}

\newcommand{\chainCont}[7]{
    \begin{tikzpicture}[scale=1.15,baseline={([yshift=-1ex]current bounding box.center)}]
        \def\r{8pt};
        \def\l{3.5};
        \def\sp{0.16};
        \coordinate (a1) at (0,0);
        \coordinate (a2) at (\l/2,0);
        \coordinate (a3) at (0,\l/2);
        \coordinate (a4) at (\l/2,\l/2);
    
        \clip (-0.55,\l/4-0.2) rectangle (3*\l/2+0.55,\l/2+0.1);
        
        \middleCoo{a1}{a2}{m12}{m12-1}{m12-2}{\sp};
        \middleCoo{a1}{a3}{m13}{m13-1}{m13-3}{\sp};
        \middleCoo{a2}{a4}{m24}{m24-2}{m24-4}{\sp};
        \middleCoo{a3}{a4}{m34}{m34-3}{m34-4}{\sp};
        \barycentreSq{a1}{a2}{a3}{a4}{b1234}{b1234-1}{b1234-2}{b1234-3}{b1234-4};
        
        \draw[obj1, ->-=0.6, shorten <= 7pt] ($(b1234)+(0,0)$) -- ($(m34)+(0,0)$);
        \draw[obj1, ->-=0.6, shorten <= 7pt] ($(b1234)+(\l,0)$) -- ($(m34)+(\l,0)$);
        \draw[obj2, ->-=0.6, shorten <= 7pt] ($(b1234)+(\l/2,0)$) -- ($(m34)+(\l/2,0)$);
        \foreach \x in {0,\l/2,\l}{
            \draw[mor] ($(m34-3)+(\x,0)$) to[out=-90, in=-90, looseness=1.5] ($(m34-4)+(\x,0)$);
            \draw[mod, rounded corners=\r] 
                ($(m13-3)+(\x,0)$) -- ($(b1234-3)+(\x,0)$) -- ($(m34-3)+(\x,0)$)
                ($(m24-4)+(\x,0)$) -- ($(b1234-4)+(\x,0)$) -- ($(m34-4)+(\x,0)$);
        }
        \draw[mod, dash pattern= on 0.2pt off 2.3pt] 
            (m13-3) -- ($(m13-3)+(-0.3,0)$)
            ($(m24-4)+(\l,0)$) -- ($(m24-4)+(\l+0.3,0)$);
        \node[yshift=1pt] at (\l/4,\l/4) {${\sss #5}$};
        \node[yshift=1pt] at (\l/4+\l/2,\l/4) {${\sss #6}$};
        \node[yshift=1pt] at (\l/4+\l,\l/4) {${\sss #7}$};
        \node[above] at (m13-3) {${\sss #1}$};
        \node[above] at ($(m13-3)+(\l/2,0)$) {${\sss #2}$};
        \node[above] at ($(m13-3)+(\l,0)$) {${\sss #3}$};
        \node[above] at ($(m13-3)+(3*\l/2,0)$) {${\sss #4}$};
        \node[yshift=-3.5pt] at ($(m34)+(0,0)$) {${\sss \morI}$};
        \node[yshift=-3.5pt] at ($(m34)+(\l/2,0)$) {${\sss \morII}$};
        \node[yshift=-3.5pt] at ($(m34)+(\l,0)$) {${\sss \morIII}$};
    \end{tikzpicture}
}

\newcommand{\MPOgate}[0]{
    \begin{tikzpicture}[scale=1.15,baseline={([yshift=-1ex]current bounding box.center)}]
        \def\r{8pt};
        \def\l{3.5};
        \def\sp{0.16};
        \coordinate (a1) at (0,0);
        \coordinate (a2) at (\l/2,0);
        \coordinate (a3) at (0,\l/2);
        \coordinate (a4) at (\l/2,\l/2);
    
        \clip (0.3,-.1) rectangle (2.75,1.85);
        
        \middleCoo{a1}{a2}{m12}{m12-1}{m12-2}{\sp};
        \middleCoo{a1}{a3}{m13}{m13-1}{m13-3}{\sp};
        \middleCoo{a2}{a4}{m24}{m24-2}{m24-4}{\sp};
        \middleCoo{a3}{a4}{m34}{m34-3}{m34-4}{\sp};
        \barycentreSq{a1}{a2}{a3}{a4}{b1234}{b1234-1}{b1234-2}{b1234-3}{b1234-4};

        \draw[obj2, rounded corners=2.5*\r, ->-=.52] ($(m12)+(3*\l/8,0)$) -- ($(b1234)+(3*\l/8,0)$) --node[mask point, pos=.5, minimum height=2pt](mp1){} (b1234) -- (m34);
        \clipAroundNode{mp1}{
            \draw[obj1, rounded corners=2.5*\r, ->-=.78] (m12) -- (b1234) -- ($(b1234)+(3*\l/8,0)$) -- ($(m34)+(3*\l/8,0)$);
        }
        
        \draw[mor] (m12-1) to[out=90, in=90, looseness=1.5] (m12-2);
        \draw[mor] (m34-3) to[out=-90, in=-90, looseness=1.5] (m34-4);
        \draw[mor] ($(m12-1)+(3*\l/8,0)$) to[out=90, in=90, looseness=1.5] ($(m12-2)+(3*\l/8,0)$);
        \draw[mor] ($(m34-3)+(3*\l/8,0)$) to[out=-90, in=-90, looseness=1.5] ($(m34-4)+(3*\l/8,0)$);
        
        \draw[mod, rounded corners=1.5*\r] (m34-4) -- (b1234-4) -- ($(b1234-3)+(3*\l/8,0)$) -- ($(m34-3)+(3*\l/8,0)$);
        \draw[mod, rounded corners=1.5*\r] (m12-2) -- (b1234-2) -- ($(b1234-1)+(3*\l/8,0)$) -- ($(m12-1)+(3*\l/8,0)$);
        \draw[mod, rounded corners=.8*\r] (m12-1) -- (b1234-1) -- (b1234) -- (b1234-3) -- (m34-3);
        \draw[mod, rounded corners=.8*\r] ($(m12-2)+(3*\l/8,0)$) -- ($(b1234-2)+(3*\l/8,0)$) -- ($(b1234)+(3*\l/8,0)$) -- ($(b1234-4)+(3*\l/8,0)$) -- ($(m34-4)+(3*\l/8,0)$);
        \node[yshift=-9pt, xshift=-4pt, fill=white, rectangle, inner sep=2pt] at ($(b1234)+(3*\l/8,0)$) {${\sss X}$};
        \node[yshift=-9pt, xshift=4pt, fill=white, rectangle, inner sep=2pt] at (b1234) {${\sss Y}$};
        
        \node[xshift=-12pt] at (b1234) {${\sss M_1}$};
        \node[xshift=12pt] at ($(b1234)+(3*\l/8,0)$) {${\sss N_2}$};
        \node[yshift=-16pt] at ($(b1234)+(3*\l/16,0)$) {${\sss M_2}$};
        \node[yshift=16pt] at ($(b1234)+(3*\l/16,0)$) {${\sss N_1}$};
    \end{tikzpicture}
}

\newcommand{\PEPS}[5]{
    \def\sty{#1}
    \def\morI{#2}
    \def\morII{#3}
    \def\morIII{#4}
    \def\morIV{#5}
    \PEPScont
}

\newcommand{\PEPScont}[7]{
	\begin{tikzpicture}[scale=1.15,baseline={([yshift=-.5ex]current bounding box.center)}]
    	\def\r{8pt};
    	\def\l{3.5};
    	\def\sp{0.08};
    	
    	\ifthenelse{#7 = 1 \OR #7 = 3}{\def\sgn{1}}{}
        \ifthenelse{#7 = 2}{\def\sgn{-1}}{}
    	
        \coordinate (a1) at (0,0);
        \coordinate (a2) at (\l,0);
        \coordinate (a3) at (\l/2,{\sgn*sqrt(3)*\l/2});
    	    	    	
    	\clip ({(0.5-\sp)*\l*cos(60)-0.05},-\sgn*0.05) rectangle ({\l-(0.5-\sp)*\l*cos(60)+0.05},{\sgn*(\l-(0.5+\sp)*\l*sin(60))+\sgn*0.05});
    	
    	\middleCoo{a1}{a2}{m12}{m12-1}{m12-2}{\sp};
    	\middleCoo{a2}{a3}{m23}{m23-2}{m23-3}{\sp};
    	\middleCoo{a1}{a3}{m13}{m13-1}{m13-3}{\sp};
    	\barycentre{a1}{a2}{a3}{b123}{b123-1}{b123-2}{b123-3}{\sp};
    	
    	\ifthenelse{#7 = 1}{
            \draw[obj,->-=0.77] (m12) --node[pos=0.45,fill=white,rectangle,inner sep=1pt] {${\sss #6}$} (b123);
            \draw[obj,->-=0.71] (b123) --node[pos=0.38,sloped,fill=white,rectangle,inner sep=0pt] {${\sss #4}$} (m13);
            \draw[obj,->-=0.71] (b123) --node[pos=0.38,sloped,fill=white,rectangle,inner sep=0pt] {${\sss #5}$} (m23);
        }{}
    	\ifthenelse{#7 = 3}{
            \draw[obj2,->-=0.7] (b123) --node[pos=0.38,fill=white,rectangle,inner sep=1pt,text=black] {${\sss #6}$} (m12);
            \draw[obj2,->-=0.8] (m13) --node[pos=0.46,sloped,fill=white,rectangle,inner sep=0pt,text=black] {${\sss #4}$} (b123);
            \draw[obj2,->-=0.8] (m23) --node[pos=0.46,sloped,fill=white,rectangle,inner sep=0pt,text=black] {${\sss #5}$} (b123);
        }{}
        \ifthenelse{#7 = 2}{
            \draw[obj,->-=0.7] (b123) --node[pos=0.38,fill=white,rectangle,inner sep=1pt] {${\sss #6}$} (m12);
            \draw[obj,->-=0.77] (m13) --node[pos=0.44,sloped,fill=white,rectangle,inner sep=0pt] {${\sss #4}$} (b123);
            \draw[obj,->-=0.77] (m23) --node[pos=0.44,sloped,fill=white,rectangle,inner sep=0pt] {${\sss #5}$} (b123);
        }{}
        
        \draw[mor] 
            (m12-1) to[out=\sgn*90, in=\sgn*90, looseness=1.5] (m12-2)
            (m13-1) to[out=-\sgn*30, in=-\sgn*30, looseness=1.5] (m13-3)
            (m23-2) to[out=-\sgn*150, in=-\sgn*150, looseness=1.5] (m23-3);
        \node[mor, circle, inner sep=4pt] at (b123) {};
        
        \draw[rounded corners=\r, \sty] 
            (m12-1) -- (b123-1) -- (m13-1)
            (m12-2) -- (b123-2) -- (m23-2)
            (m23-3) -- (b123-3) -- (m13-3);
        \node[yshift=-\sgn*5pt, xshift=-8pt] at (b123-1) {${\sss #1}$};
        \node[yshift=-\sgn*5pt, xshift=8pt] at (b123-2) {${\sss #2}$};
        \node[yshift=\sgn*9pt] at (b123-3) {${\sss #3}$}; 
        \node[] at (b123) {${\sss \morII}$};
        \node[yshift=\sgn*3.5pt] at (m12) {${\sss \morIII}$};
        \node[yshift=-\sgn*1.5pt, xshift=2.5pt] at (m13) {${\sss \morI}$};
        \node[yshift=-\sgn*1.5pt, xshift=-2.5pt] at (m23) {${\sss \morIV}$};
	\end{tikzpicture}
}

\newcommand{\PEPSgate}[0]{
    \begin{tikzpicture}[scale=1.15,baseline={([yshift=-1ex]current bounding box.center)}]
        \def\r{8pt};
        \def\l{3.5};
        \def\sp{0.16};
        \coordinate (a1) at (0,0);
        \coordinate (a2) at (\l/2,0);
        \coordinate (a3) at (0,\l/2);
        \coordinate (a4) at (\l/2,\l/2);
    
        \clip (-.2,-.1) rectangle (1.9,1.85);
        
        \middleCoo{a1}{a2}{m12}{m12-1}{m12-2}{\sp};
        \middleCoo{a1}{a3}{m13}{m13-1}{m13-3}{\sp};
        \middleCoo{a2}{a4}{m24}{m24-2}{m24-4}{\sp};
        \middleCoo{a3}{a4}{m34}{m34-3}{m34-4}{\sp};
        \barycentreSq{a1}{a2}{a3}{a4}{b1234}{b1234-1}{b1234-2}{b1234-3}{b1234-4};

        \draw[obj2, ->-=.7] (b1234) -- node[pos=0.38,fill=white,rectangle,inner sep=1pt,text=black] {${\sss X}$}(m34);
        \draw[obj2, rounded corners=2.7*\r, ->-=.75] ($(m12)+(3*\l/16,0)$) -- ($(b1234)+(3*\l/16,0)$) -- (b1234);
        \draw[obj2, rounded corners=2.7*\r, ->-=.75] ($(m12)+(-3*\l/16,0)$) -- ($(b1234)+(-3*\l/16,0)$) -- (b1234); 
        \node[fill=white, rectangle, inner sep=1pt, xshift=20pt, yshift=-12pt] at (b1234) {${\sss X''}$};
        \node[fill=white, rectangle, inner sep=1pt, xshift=-19pt, yshift=-12pt] at (b1234) {${\sss X'}$};
    
        \draw[mor] (m34-3) to[out=-90, in=-90, looseness=1.5] (m34-4);
        \draw[mor] ($(m12-1)+(-3*\l/16,0)$) to[out=90, in=90, looseness=1.5] ($(m12-2)+(-3*\l/16,0)$);
        \draw[mor] ($(m12-1)+(3*\l/16,0)$) to[out=90, in=90, looseness=1.5] ($(m12-2)+(3*\l/16,0)$);
        \node[mor, circle, inner sep=4pt] at (b1234) {};
        \node at (b1234) {${\sss i}$};

        \draw[mod, rounded corners=1.5*\r] ($(m12-2)+(-3*\l/16,0)$) -- ($(b1234-2)+(-3*\l/16,0)$) -- ($(b1234-1)+(3*\l/16,0)$) -- ($(m12-1)+(3*\l/16,0)$);
        \draw[mod, rounded corners=1.4*\r] ($(m12-1)+(-3*\l/16,0)$) -- ($(b1234-1)+(-3*\l/16,0)$) -- (b1234-3) -- (m34-3);
       \draw[mod, rounded corners=1.4*\r] ($(m12-2)+(3*\l/16,0)$) -- ($(b1234-2)+(3*\l/16,0)$) -- (b1234-4) -- (m34-4); 

        \node[yshift=-16pt] at (b1234) {${\sss N}$};
        \node[xshift=19pt, yshift=10pt] at (b1234) {${\sss O}$};
        \node[xshift=-19pt, yshift=10pt] at (b1234) {${\sss M}$};
    \end{tikzpicture}
}

\newcommand{\gate}[1]{
    \begin{tikzpicture}[scale=1.15,baseline={([yshift=-1ex]current bounding box.center)}]
        \def\l{.6};
        \def\h{.3};
        \def\s{1.2*\l};
        \coordinate (a) at (0,0);
        \ifthenelse{#1=0}{
            \draw[obj1, ->-=.7] ($(a)+(-\l/2,-\s)$) -- ($(a)+(-\l/2,0)$);
            \draw[obj2, ->-=.7] ($(a)+(-\l/2,0)$) -- ($(a)+(-\l/2,\s)$);
            \draw[obj2, ->-=.5] ($(a)+(\l/2,-\s)$) -- ($(a)+(\l/2,0)$);
            \draw[obj1, ->-=.7] ($(a)+(\l/2,0)$) -- ($(a)+(\l/2,\s)$);
            \blob{0}{0}{\l}{\h}{\omega}{1};
            \node[below, yshift=1pt, fill=white, rectangle, inner sep=1pt] at ($(a)+(-\l/2,-.73*\l)$) {${\sss Y}$};
        }{}
        \ifthenelse{#1=1}{
            \draw[obj1, ->-=.7] ($(a)+(-\l/2,-\s)$) -- ($(a)+(-\l/2,0)$);
            \draw[obj2, ->-=.7] ($(a)+(-\l/2,0)$) -- ($(a)+(-\l/2,\s)$);
            \draw[obj2, ->-=.7] ($(a)+(\l/2,-\s)$) -- ($(a)+(\l/2,0)$);
            \draw[obj1, ->-=.7] ($(a)+(\l/2,0)$) -- ($(a)+(\l/2,\s)$);
            \blob{0}{0}{\l}{\h}{\omega}{1};
            \node[below, yshift=1pt, fill=white, rectangle, inner sep=1pt] at ($(a)+(-\l/2,-.73*\l)$) {${\sss Y}$};
            \node[below, yshift=1pt, fill=white, rectangle, inner sep=1pt] at ($(a)+(\l/2,-.73*\l)$) {${\sss X}$};
        }{}
        \ifthenelse{#1=2 \OR #1=3}{
            \draw[obj2, ->-=.7] ($(a)+(-\l/2,-\s)$) -- ($(a)+(-\l/2,0)$);
            \draw[obj2, ->-=.88] ($(a)+(-\l/2,0)$) -- ($(a)+(-\l/2,\s)$);
            \draw[obj2, ->-=.7] ($(a)+(\l/2,-\s)$) -- ($(a)+(\l/2,0)$);
            \draw[obj3, ->-=.7] ($(a)+(\l/2,0)$) -- ($(a)+(\l/2,\s)$);
            \blob{0}{0}{\l}{\h}{{}^{\fr F} \! F}{2};
            \node[below, yshift=2pt, fill=white, rectangle, inner sep=1pt] at ($(a)+(-\l/2,-.73*\l)$) {${\sss X'}$};
            \node[below, yshift=2pt, fill=white, rectangle, inner sep=1pt] at ($(a)+(\l/2,-.73*\l)$) {${\sss X''}$};
            \ifthenelse{#1=2}{
                \node[above, yshift=-1pt] at ($(a)+(\l/2,\s)$) {${\sss \la i|}$};
            }{
                \node[above, yshift=-1pt] at ($(a)+(\l/2,\s)$) {${\sss \phantom{|i\ra}}$};
            }
            \node[above, yshift=-1pt, fill=white, rectangle, inner sep=1pt] at ($(a)+(-\l/2,.44*\l)$) {${\sss X}$};
        }{}
        \ifthenelse{#1=6}{
            \draw[obj2, ->-=.5] ($(a)+(-\l/2,-\s)$) -- ($(a)+(-\l/2,0)$);
            \draw[obj2, ->-=.7] ($(a)+(-\l/2,0)$) -- ($(a)+(-\l/2,\s)$);
            \draw[obj2, ->-=.5] ($(a)+(\l/2,-\s)$) -- ($(a)+(\l/2,0)$);
            \draw[obj3, ->-=.7] ($(a)+(\l/2,0)$) -- ($(a)+(\l/2,\s)$);
            \blob{0}{0}{\l}{\h}{{}^{\fr F} \! F}{2};
            \node[above, yshift=-1pt] at ($(a)+(\l/2,\s)$) {${\sss \,\, \la X''\! X' \! X,i|}$};
        }{}
        \ifthenelse{#1=4}{
            \draw[obj2, ->-=.7] ($(a)+(-\l/2,-\s)$) -- ($(a)+(-\l/2,0)$);
            \draw[obj2, ->-=.7] ($(a)+(-\l/2,0)$) -- ($(a)+(-\l/2,\s)$);
            \draw[obj3, ->-=.5] ($(a)+(\l/2,-\s)$) -- ($(a)+(\l/2,0)$);
            \draw[obj2, ->-=.7] ($(a)+(\l/2,0)$) -- ($(a)+(\l/2,\s)$);
            \blob{0}{0}{\l}{\h}{{}^{\fr F} \! \bar F}{2};
            \node[below, yshift=2pt, fill=white, rectangle, inner sep=1pt] at ($(a)+(-\l/2,-.73*\l)$) {${\sss \mathbb 1}$};
            \node[below, yshift=2pt] at ($(a)+(\l/2,-\s)$) {${\sss |+\ra}$};
            \node[below, yshift=2pt] at ($(a)+(-\l/2,-\s)$) {${\sss |\mathbb 1\ra}$};
        }{}
        \ifthenelse{#1=5}{
            \draw[obj2, ->-=.7] ($(a)+(-\l/2,-\s)$) -- ($(a)+(-\l/2,0)$);
            \draw[obj2, ->-=.88] ($(a)+(-\l/2,0)$) -- ($(a)+(-\l/2,\s)$);
            \draw[obj3, ->-=.5] ($(a)+(\l/2,-\s)$) -- ($(a)+(\l/2,0)$);
            \draw[obj2, ->-=.88] ($(a)+(\l/2,0)$) -- ($(a)+(\l/2,\s)$);
            \blob{0}{0}{\l}{\h}{{}^{\fr F} \! \bar F}{2};
            \node[below, yshift=2pt, fill=white, rectangle, inner sep=1pt] at ($(a)+(-\l/2,-.73*\l)$) {${\sss \mathbb 1}$};
            \node[above, yshift=-1pt, fill=white, rectangle, inner sep=1pt] at ($(a)+(-\l/2,.44*\l)$) {${\sss X}$};
            \node[above, yshift=-1pt, fill=white, rectangle, inner sep=1pt] at ($(a)+(\l/2,.44*\l)$) {${\sss \bar X}$};
            \node[below, yshift=2pt] at ($(a)+(\l/2,-\s)$) {${\sss |1\ra}$};
            \node[below, yshift=2pt] at ($(a)+(-\l/2,-\s)$) {${\sss |\mathbb 1\ra}$};
        }{}
        \ifthenelse{#1=7}{
            \draw[obj2, ->-=.7] ($(a)+(-\l/2,0)$) -- ($(a)+(-\l/2,\s)$);
            \draw[obj2, ->-=.7] ($(a)+(\l/2,0)$) -- ($(a)+(\l/2,\s)$);
            \blob{0}{0}{\l}{1.4*\h}{\mc C^{(i)}_h}{2};
        }{}
        \ifthenelse{#1=8}{
            \draw[obj2, ->-=.7] ($(a)+(-\l/2,-\s)$) -- ($(a)+(-\l/2,0)$);
            \draw[obj2, ->-=.7] ($(a)+(-\l/2,0)$) --node[left, text=black, xshift=2pt](){${\sss C^{(i)}_h}$} ($(a)+(-\l/2,\s)$);
            \draw[obj3, ->-=.5] ($(a)+(\l/2,-\s)$) -- ($(a)+(\l/2,0)$);
            \draw[obj2, ->-=.7] ($(a)+(\l/2,0)$) --node[left, text=black, xshift=2pt](){${\sss \bar C^{(i)}_h}$} ($(a)+(\l/2,\s)$);
            \blob{0}{0}{\l}{\h}{{}^{\fr F} \! \bar F}{2};
            \node[below, yshift=2pt, fill=white, rectangle, inner sep=1pt] at ($(a)+(-\l/2,-.73*\l)$) {${\sss \mathbb 1}$};
            \node[below, yshift=2pt] at ($(a)+(\l/2,-\s)$) {${\sss |1\ra}$};
            \node[below, yshift=2pt] at ($(a)+(-\l/2,-\s)$) {${\sss |\mathbb 1\ra}$};
        }{}
    \end{tikzpicture}
}

\newcommand{\circuitClosed}[0]{
    \begin{tikzpicture}[scale=1.15,baseline={([yshift=-1ex]current bounding box.center)}]
        \def\l{.6};
        \def\h{.3};
        \def\s{1.2*\l};
        \draw[obj3, ->-=.5] ($(0,0)+(\l/2,-\s)$) -- ($(0,0)+(\l/2,0)$);
        \draw[obj2, ->-=.55] ($(0,0)+(\l/2,0)$) to[out=90, in=180, looseness=1] ($(0,0)+(2*\l,4.5*\h+.5*\s)$);
        \draw[obj2, ->-=.7] ($(0,0)+(-\l/2,-\s)$) -- ($(0,0)+(-\l/2,0)$);
        \draw[obj2, ->-=.6] ($(0,0)+(-\l/2,0)$) -- ($(0,0)+(-\l/2,2*\h)$);
        \draw[obj1, ->-=.4] ($(0,0)+(-\l/2,2*\h)$) --node[pos=.6,fill=white, rectangle, inner sep=-2pt](){${\sss Y_{L-\frac{1}{2}}}$} ($(0,0)+(-\l/2,11*\h+\s)$);
        \draw[obj1, ->-=.7] ($(-\l,0)+(-\l/2,-\s)$) --node[pos=.35, fill=white, rectangle, inner sep=-2pt](){${\sss Y_{L-\frac{1}{2}}}$} ($(-\l,0)+(-\l/2,2*\h)$);
        \draw[obj2] ($(-\l,0)+(-\l/2,2*\h)$) -- ($(-\l,0)+(-\l/2,3*\h)$);    
        \draw[obj2] ($(-2*\l,0)+(-\l/2,5*\h)$) -- ($(-2*\l,0)+(-\l/2,4*\h)$);
        \draw[obj1, ->-=.4] ($(-2*\l,0)+(-\l/2,5*\h)$) --node[pos=.6, ontop](){${\sss Y_\frac{3}{2}}$} ($(-2*\l,0)+(-\l/2,11*\h+\s)$);
        \draw[obj1, ->-=.6] ($(-3*\l,0)+(-\l/2,-\s)$) --node[pos=.3,ontop](){${\sss Y_\frac{3}{2}}$} ($(-3*\l,0)+(-\l/2,5*\h)$);
        \draw[obj2, ->-=.6] ($(-3*\l,0)+(-\l/2,5*\h)$) -- ($(-3*\l,0)+(-\l/2,7*\h)$);
        \draw[obj1, ->-=.4] ($(-3*\l,0)+(-\l/2,7*\h)$) --node[pos=.6,ontop](){${\sss Y_\frac{1}{2}}$} ($(-3*\l,0)+(-\l/2,11*\h+\s)$);
        \draw[obj1, ->-=.65] ($(-4*\l,0)+(-\l/2,-\s)$) --node[pos=.35,ontop](){${\sss Y_\frac{1}{2}}$} ($(-4*\l,0)+(-\l/2,7*\h)$);
        \draw[obj2, ->-=.6] ($(-4*\l,0)+(-\l/2,7*\h)$) -- ($(-4*\l,0)+(-\l/2,9*\h)$);
        \draw[obj3, ->-=.6] ($(-4*\l,0)+(-\l/2,9*\h)$) -- ($(-4*\l,0)+(-\l/2,11*\h+\s)$);
        \draw[obj2, ->-=.6] ($(-5*\l,0)+(-\l/2,9*\h)$) -- ($(-5*\l,0)+(-\l/2,11*\h)$);
        \draw[obj3, ->-=.7] ($(-5*\l,0)+(-\l/2,11*\h)$) -- ($(-5*\l,0)+(-\l/2,11*\h+\s)$);
        \draw[obj2, ->-=.7] ($(-6*\l,0)+(-\l/2,11*\h)$) -- ($(-6*\l,0)+(-\l/2,11*\h+\s)$);
        \draw[obj2, ->-=.68] ($(-5*\l,0)+(-\l/2,-\s)$) -- ($(-5*\l,0)+(-\l/2,9*\h)$);
        \draw[obj2, ->-=.55] ($(-6*\l,0)+(-2*\l,4.5*\h+.5*\s)$)to[out=0, in=-90, looseness=1] (-6*\l-.5*\l,11*\h);

        \blob{0}{0}{\l}{\h}{{}^\fr F \! \bar F}{2};
        \blob{-\l}{2*\h}{\l}{\h}{\omega}{1};
        \blob{-3*\l}{5*\h}{\l}{\h}{\omega}{1};
        \blob{-4*\l}{7*\h}{\l}{\h}{\omega}{1};
        \blob{-5*\l}{9*\h}{\l}{\h}{{}^\fr F \! F}{2};
        \blob{-6*\l}{11*\h}{\l}{\h}{{}^\fr F \! F}{2};
        
        \node[below, fill=white, rectangle, inner sep=1pt] at (-\l/2,-.68*\l) {${\sss \mathbb 1}$}; 
        \node[below, yshift=2pt] at ($(0,0)+(\l/2,-\s)$) {${\sss |+\ra}$};
        \node[below, yshift=2pt] at ($(0,0)+(\l/2,-\s)$) {${\sss |+\ra}$};
        \node[below, yshift=2pt] at ($(0,0)+(-\l/2,-\s)$) {${\sss |\mathbb 1\ra}$};
        \node at (-2*\l,3.5*\h) {${\sss \ldots}$};
        \node[color=BrickRed] at ($(-\l,0)+(-\l/2,3.5*\h)$) {\rotatebox[origin=c]{90}{${\sss \ldots}$}};
        \node[color=BrickRed] at ($(-3*\l,0)+(\l/2,3.5*\h)$) {\rotatebox[origin=c]{90}{${\sss \ldots}$}};
    \end{tikzpicture}
}

\newcommand{\circuitConstant}[0]{
    \begin{tikzpicture}[scale=1.15,baseline={([yshift=-1ex]current bounding box.center)}]
        \def\l{.6};
        \def\h{.6};
        \def\s{1.2*\l};   

        \draw[obj2,->-=.7] (-\l-\l/2,-\h-\s) --node[pos=.3, ontop](){${\sss A}$} (-\l-\l/2,\h);
        \draw[obj2] (4*\l+\l/2,-\h) -- (4*\l+\l/2,-\h+\h/2);
        \draw[obj2] (6*\l-\l/2,\h) -- (6*\l-\l/2,\h-\h/2);
        
        \node[color=BrickRed] at (4*\l+\l/2,-\h+3*\h/4) {\rotatebox[origin=c]{90}{${\sss \ldots}$}};
        \node[color=BrickRed] at (5*\l+\l/2,\h-3*\h/4) {\rotatebox[origin=c]{90}{${\sss \ldots}$}};
        \node at (5*\l,0) {${\sss \ldots}$};
        
        \draw[obj2, ->-=.8] (8*\l+\l/2,-\h) to[out=90, in=180, looseness=1] node[ontop,pos=.4](){${\sss \bar X_L}$} (9*\l,\h/2);
        \draw[obj2, ->-=.5] (-3*\l,\h/2) to[out=0, in=-90, looseness=1] (-2*\l-\l/2,2*\h);
        \draw[obj2,->-=.73] (\l+\l/2,-\h) --node[ontop,pos=.38](){${\sss \bar{X}_1}$} (\l+\l/2,\h);
        \draw[obj2, ->-=.6] (-\l-\l/2,\h) --node[pos=.45,xshift=-7pt,text=black](){${\sss X_1'}$} (-\l-\l/2,2*\h);
        \draw[obj3,->-=.7] (-2*\l+\l/2,2*\h) -- (-2*\l+\l/2,2*\h+\s);
        \node[above, yshift=-2pt, xshift=1pt] at (-2*\l+\l/2,2*\h+\s) {${\sss \la j|}$};
        
        \draw[obj2, ->-=.8] (-2*\l-\l/2,2*\h) to[out=90,in=-90,looseness=1] node[pos=.4,ontop](){${\sss B^{(0)}_{g}}$} (-\l-\l/2,6*\h);
        \draw[obj2,->-=.7] (2*\l-\l/2,\h) to[out=90,in=-90,looseness=1] node[pos=.4,ontop](){${\sss B^{(0)}_{2,g_2}}$} (2*\l,6*\h);
        \draw[obj2,->-=.7] (6*\l-\l/2,\h) to[out=90,in=-90,looseness=1] node[ontop,pos=.4](){${\sss B^{(0)}_{L,g_L}}$} (6*\l,7*\h);
        
        \foreach \x/\i/\n in {0/\frac{1}{2}/1,3*\l/\frac{3}{2}/2,7*\l/L-\frac{1}{2}/L}{
            \draw[obj1,->-=.5] (\x+\l/2,0) to[out=90, in=-90, in looseness=1, out looseness=2] node[pos=.25, ontop](){${\sss Y_{\i}}$} (\x+\l/2-\l,2*\h+\s) -- (\x+\l/2-\l,5*\h);
            \draw[obj1, ->-=.7] (\x-\l/2,-\h-\s) --node[pos=.3, ontop](){${\sss Y_{\i}}$} (\x-\l/2,0);
            \draw[obj2, ->-=.6] (\x+\l/2,-\h) --node[pos=.45,xshift=-7pt,text=black](){${\sss X_\n}$} (\x+\l/2,0);
            \draw[obj2, ->-=.6] (\x-\l/2,0) -- (\x-\l/2,\h);
            \draw[obj3, ->-=.75] (\x-\l/2,\h) -- (\x-\l/2,\h+.7*\s);
            \draw[obj2, ->-=.7] (\x+\l-\l/2,-\h-\s) --node[pos=.3,ontop](){${\sss \mathbb 1}$} (\x+\l-\l/2,-\h);
            \draw[obj3, ->-=.5] (\x+\l+\l/2,-\h-\s) -- (\x+\l+\l/2,-\h);
            \blob{\x}{0}{\l}{\h/2}{\omega}{1};
            \blob{\x+\l}{-\h}{\l}{\h/2}{{}^\fr F \! \bar F}{2};
            \blob{\x-\l}{\h}{\l}{\h/2}{{}^\fr F \! F}{2};
            \node[below, yshift=2pt] at (\x+\l-\l/2,-\h-\s) {${\sss |\mathbb 1 \ra}$};
            \node[below, yshift=2pt] at (\x+\l+\l/2,-\h-\s) {${\sss |\mathbb + \ra}$};
            \node[above, yshift=-2pt] at (\x-\l/2,\h+.7*\s) {${\sss \la i_\n |}$};
        }
        \blob{-2*\l}{2*\h}{\l}{\h/2}{{}^\fr F \! F}{2};

        \draw[obj2, ->-=.6] (\l/2,4*\h) --node[xshift=-9pt,pos=.4,text=black](){${\sss C^{(0)}_{h_2}}$} (\l/2,5*\h);
        \draw[obj2, ->-=.6] (3*\l+\l/2,4*\h) --node[xshift=-9pt,pos=.4,text=black](){${\sss C^{(0)}_{h_3}}$} (3*\l+\l/2,5*\h);
        \draw[obj2, ->-=.6] (3*\l-\l/2,5*\h) -- (3*\l-\l/2,6*\h);
        \draw[obj2, ->-=.6] (2*\l,6*\h) -- (2*\l,7*\h);
        \draw[obj2, ->-=.7] (\l+\l/2,4*\h) --node[pos=.4,ontop](){${\sss \bar C^{(0)}_{h_2}}$} (\l+\l/2,7*\h);
        \draw[obj3, ->-=.75] (3*\l-\l/2,6*\h) -- (3*\l-\l/2,6*\h+.7*\s);
        \draw[obj3, ->-=.7] (2*\l,7*\h) -- (2*\l,7*\h+\s);
        \draw[obj2, ->-=.7] (2*\l-\l/2,7*\h) --node[pos=.6,xshift=-11pt,text=black](){${\sss B^{(1)}_{2,k_2}}$} (2*\l-\l/2,7*\h+\s);
        \draw[obj2,->-=.6] (-\l/2,5*\h) -- (-\l/2,6*\h);
        \draw[obj2, ->-=.8] (-\l-\l/2,6*\h) --node[ontop,pos=.4](){${\sss B'^{(0)}_{g'}}$} (-\l-\l/2,7*\h+\s);
        \draw[obj3, ->-=.75] (-\l+\l/2,6*\h) -- (-\l+\l/2,6*\h+.7*\s);
        \draw[obj1,->-=.8] (\l/2,5*\h) to[out=90, in=-90, in looseness=1, out looseness=2] node[pos=.3, ontop](){${\sss Y_\frac{1}{2}}$} (-\l/2,7*\h+\s);
        \draw[obj1,->-=.8] (3*\l+\l/2,5*\h) to[out=90, in=-90, in looseness=1, out looseness=2] node[pos=.3, ontop](){${\sss Y_\frac{3}{2}}$} (3*\l-\l/2,7*\h+\s);
        \draw[obj1] (7*\l-\l/2,4.8*\h) -- (7*\l-\l/2,7*\h+\s);
        \draw[obj3, ->-=.7] (6*\l,7*\h) -- (6*\l,7*\h+\s);
        \draw[obj2] (6*\l-\l/2,7*\h) -- (6*\l-\l/2,7*\h-\h/2);
        \draw[obj2, ->-=.7] (6*\l-\l/2,7*\h) --node[pos=.6,xshift=-11pt,text=black](){${\sss B^{(1)}_{L,k_L}}$} (6*\l-\l/2,7*\h+\s);
        \draw[obj2] (4*\l+\l/2,4*\h) -- (4*\l+\l/2,4*\h+\h/2);
        \node[above, yshift=-2pt] at (6*\l,7*\h+\s) {${\sss \la j'_L|}$};
        \node[above, yshift=-2pt] at (2*\l,7*\h+\s) {${\sss \la j'_2|}$};
        \node[above, yshift=-2pt] at (-\l+\l/2,6*\h+.7*\s) {${\sss \la j'_1 |}$};
        \node[above, yshift=-2pt] at (2*\l+\l/2,6*\h+.7*\s) {${\sss \la i'_2 |}$};
        
        \blob{\l}{3.88*\h}{\l}{1.4*\h/2}{\mc C^{(0)}_{h_2}}{2}; 
        \blob{4*\l}{3.88*\h}{\l}{1.4*\h/2}{\mc C^{(0)}_{h_3}}{2};  
        \blob{0}{5*\h}{\l}{\h/2}{\omega}{1};   
        \blob{3*\l}{5*\h}{\l}{\h/2}{\omega}{1}; 
        \blob{2.25*\l}{6*\h}{\l/2}{\h/2}{{}^\fr F \! F}{2};
        \blob{1.75*\l}{7*\h}{\l/2}{\h/2}{{}^\fr F \! F}{2};
        \blob{5.75*\l}{7*\h}{\l/2}{\h/2}{{}^\fr F \! F}{2};
        \blob{-\l}{6*\h}{\l}{\h/2}{{}^\fr F \! F}{2};

        \node[color=BrickRed] at (4*\l+\l/2,4.75*\h) {\rotatebox[origin=c]{90}{${\sss \ldots}$}};
        \node[color=BrickRed] at (5*\l+\l/2,6.25*\h) {\rotatebox[origin=c]{90}{${\sss \ldots}$}};
        \node at (5*\l,5.5*\h) {${\sss \ldots}$};
    \end{tikzpicture}
}

%% file: _Introduction.tex
\noindent
{\bf Introduction:} 
Dualities have long played important roles in condensed matter theory, from locating critical points \cite{PhysRev.60.252} or even completely solving models \cite{1928ZPhy...47..631J,PhysRevLett.63.322}, to mapping non-local order parameters into local ones \cite{PhysRevB.45.304,cmp/1104250747} and generating interesting patterns of entanglement \cite{PhysRevX.5.011024,PhysRevB.88.085114,PRXQuantum.4.020315}. Given two dual models, there is the expectation that their spectra can be related via an isometry \cite{Cobanera_2011}, and that a (unitary) quantum circuit can implement the duality, which is the first step towards experimental realization on a quantum device. However, dualities are typically non-invertible on a fixed Hilbert space; only when addressing the non-trivial mapping of boundary conditions and charge sectors can such an isometry be constructed. Recently, a general systematic approach to dualities in symmetric one-dimensional quantum lattice models has been proposed \cite{PRXQuantum.4.020357}---echoing a modern viewpoint on symmetry in terms of topological defects \cite{Gaiotto:2014kfa}---whereby dual models only differ in a choice of module category over a fusion category encoding the symmetry of the model in question. In this approach, the isometries associated with compatible boundary conditions and charge sectors are realized in the form of \emph{matrix product operators} \cite{Lootens:2022avn}.

In this paper, we demonstrate that the duality operators of ref.~\cite{Lootens:2022avn} can be realized as unitary linear-depth quantum circuits when supplementing the Hilbert space with ancillary degrees of freedom that keep track of the various sectors. Measuring the ancillary degrees of freedom has the effect of projecting input and output states into definite sectors, at which point the operator boils down to the unitary circuit relating the spectra. Since generic operators are mapped to string operators upon dualizing, the linear depth of these (ordinary) quantum circuits is optimal. However, we demonstrate that constant depth can be achieved with \emph{adaptive} quantum circuits \cite{1999Natur.402..390G,v001a005,jozsa2005introduction} in the case of \emph{nilpotent dualities} defined as dualities for which the dual symmetry is encoded into a nilpotent fusion category \cite{GELAKI20081053}. This requires the introduction of an extensive amount of ancillary degrees of freedom combined with measurements and classical communication, as initially demonstrated in ref.~\cite{Tantivasadakarn:2021vel,PRXQuantum.4.020339,Bravyi:2022zcw} for certain special cases.

%% file: _Dualities.tex
\medskip\noindent
\textbf{Duality operators:}  We begin by briefly reviewing the tools necessary to address dualities in full generality \cite{PRXQuantum.4.020357,Lootens:2022avn}. Loosely speaking, a fusion category $\mc D$ consists of a collection of \emph{objects} interpreted as topological charges whose quantum dimensions are not necessarily integers. We denote representatives of isomorphism classes of simple objects via $Y_1,Y_2,\ldots \in \mc I_\mc D$ and their respective quantum dimensions via $d_{Y_1}, d_{Y_2}, \ldots \in \mathbb C$. Naturally, topological charges can be fused, which is encoded into a \emph{fusion} structure $(\otimes, \mathbb 1,F)$ consisting of a product rule $\otimes : \mc D \times \mc D \to \mc D$, a trivial topological charge $\mathbb 1$, and a collection $F$ of isomorphisms $F^{Y_1 Y_2 Y_3} : (Y_1 \otimes Y_2) \otimes Y_3 \xrightarrow{\sim} Y_1 \otimes (Y_2 \otimes Y_3)$ for every $Y_1,Y_2,Y_3 \in \mc I_\mc D$. These isomorphisms are required to satisfy a coherence relation referred to as the `pentagon axiom' \cite{etingof2016tensor}. Fusion categories play two related roles in our construction: On the one hand, a fusion category $\mc D$ furnishes an abstract algebra of local symmetric operators, a subalgebra of which specifying an equivalence class of physical models. On the other hand, fusion categories organize the symmetry operators of these models. 

Given abstract local symmetric operators building up an equivalence class of models, explicit matrix representations are obtained via a choice of module category $\mc M$ over $\mc D$, which loosely amounts to picking a family of physical degrees of freedom \cite{Lootens:2022avn}. More concretely, a (right) $\mc D$-module category $(\mc M,\cat,\F{\cat})$ consists of a collection of objects, representatives of their isomorphisms classes are denoted by $M_1,M_2,\ldots \in \mc I_\mc M$, an action $\cat : \mc M \times \mc D \to \mc M$ and a collection $\F{\cat}$ of isomorphisms $\F{\cat}^{MY_1Y_2} : (M \cat Y_1) \cat Y_2 \xrightarrow{\sim} M \cat (Y_1 \otimes Y_2)$ for every $M \in \mc I_\mc M$ and $Y_1,Y_2 \in \mc I_D$. These isomorphisms are also required to satisfy a `pentagon axiom' \cite{Ostrik2003-ao,10.1155/S1073792803205079,etingof2016tensor}. For instance, $\mc D$ is a module category over itself with $\cat = \otimes$ and $F = \F{\cat}$. Denoting by $\mc V^{M_2}_{M_1Y}:= \Hom_{\mc M}(M_1 \cat Y,M_2) \ni |M_1YM_2,i\ra$ the vector space of maps from $M_1 \cat Y$ to $M_2$, these isomorphisms boil down to collections of complex matrices:
\begin{equation}
    \F{\cat}^{M_1Y_1Y_2}_{M_2} : 
    \bigoplus_{M_3} \mc V^{M_3}_{M_1 Y_1} \otimes \mc V^{M_2}_{M_3 Y_2}
    \xrightarrow{\sim}
    \bigoplus_{Y_3} \mc H^{Y_3}_{Y_1 Y_2} \otimes \mc V^{M_2}_{M_1 Y_3} \, ,
\end{equation}
where $\mc H^{Y_3}_{Y_1 Y_2} := \Hom_{\mc D}(Y_1 \otimes Y_2,Y_3)$.
Given the data of a $\mc D$-module category, we assign to an \emph{infinite} chain a microscopic Hilbert space given by the $\mathbb C$-linear span \cite{Lootens:2022avn}
\begin{align}
    \label{eq:infHilbert}
    \mathbb C \bigg[\!\!
    \infChain{i_{\msf i-\frac{1}{2}}}{i_{\msf i+\frac{1}{2}}}{i_{\msf i+\frac{3}{2}}}{M_{\msf i-1}}{M_{\msf i}}{M_{\msf i+1}}{M_{\msf i+2}}{Y_{\msf i - \frac{1}{2}}}{Y_{\msf i+\frac{1}{2}}}{Y_{\msf i + \frac{3}{2}}} \!\!\! \bigg] 
\end{align}
over simple objects $\{M \in \mc I_\mc M\}$, $\{Y \in \mc I_\mc D\}$ and basis vectors $\{|M_\msf i Y_{\msf i+\frac{1}{2}} M_{\msf i+1},i_{\msf i+\frac{1}{2}} \ra\}$ in the corresponding hom-spaces, which may be zero. Given a collection of local symmetric operators defined with respect to such a Hilbert space, any weighted sum of these operators results in a Hamiltonian generically denoted by $\mathbb H^\mc M$.

Duality operators are provided by maps between $\mc D$-module categories that are compatible with the $\mc D$-action, independently of the specific physical models we are interested in \cite{Lootens:2022avn}. More precisely, given two $\mc D$-module categories $(\mc M, \cat ,\F{\cat})$ and $(\mc N, \catb, \F{\catb})$, a $\mc D$-module functor $(\fr F,\omega)$ between them consists of a functor $\fr F: \mc M \to \mc N$ and a collection $\omega$ of isomorphisms $\omega^{M Y} : \fr F(M \cat Y) \xrightarrow{\sim} \fr F(M) \catb Y$, for every $Y \in \mc I_\mc D$ and $M \in \mc I_\mc M$, which boil down to collections of complex matrices:
\begin{equation}
    \label{eq:omF}
    \omega^{M_1Y}_{N_2} : \bigoplus_{N_1} \mc V^{N_1}_{\fr F M_1} \otimes \mc V^{N_1}_{N_2 Y} \xrightarrow{\sim} 
    \bigoplus_{M_2}\mc V^{M_2}_{M_1 Y} \otimes \mc V^{N_2}_{\fr F M_2} \, ,    
\end{equation}
where $\mc V^{N}_{\fr F M} := \Hom_\mc N(\fr F(M),N)$.
These isomorphisms are also required to satisfy a `pentagon axiom' \cite{etingof2016tensor}. Such $\mc D$-module functors are organized into a category denoted by $\Fun_\mc D(\mc M,\mc N)$. We use $X_1,X_2, \ldots$ to denote representatives of isomorphism classes of simple objects in such categories, in which case the corresponding $\mc D$-module structure is denoted by $(\Func{X},\omF{X})$. Let us explicitly construct the duality operator associated with an object $X$ in $\Fun_\mc D(\mc M,\mc N)$. First we introduce a graphical notation for the entries of matrices $\omF{X}^{M_1Y}_{N_2}$ as defined in eq.~\eqref{eq:omF}:
\begin{align}
    \MPO{mod}{mod}{i}{j}{k}{l}{N_1}{N_2}{M_2}{M_1}{Y}{X}{1} \equiv
    &\big( \omF{X}^{M_1Y}_{N_2}\big)^{M_2,lj}_{N_1,ik} \, ,
\end{align}
where indices $i,j,k,l$ label basis vectors in the hom-spaces $\mc V^{N_1}_{X M_1}$, $\mc V^{N_2}_{X M_2}$, $\mc V^{N_2}_{N_1Y}$ and $\mc V^{M_2}_{M_1Y}$, respectively. Introducing the graphical convention
\begin{equation}
    \begin{split}
        \label{eq:unlabelledPatch}
        \splitSpace{M_1}{Y}{M_2}{}{1} &\equiv
        \sum_{i}
        \splitSpace{M_1}{Y}{M_2}{i}{1}
        |M_1YM_2, i \ra \, , 
        \\
        \splitSpace{M_1}{Y}{M_2}{}{2} &\equiv
        \sum_{i}
        \splitSpace{M_1}{Y}{M_2}{i}{2} \la M_1YM_2, i | \, , 
    \end{split}
\end{equation}
we construct the following duality operator in the form of a matrix product operator \cite{10.21468/SciPostPhys.10.3.053}:
\begin{equation}
    \label{eq:symMPO}
    \sum_{\substack{\{Y \in \mc I_\mc D \} \\ \{M \in \mc I_\mc M\} \\ \{N \in \mc I_\mc N\}}}
    \intertwiner{1} \, .
\end{equation} 
It follows from the pentagon axiom satisfied by isomorphisms $\omF{X}$ that this duality operator transmutes a symmetric Hamiltonian $\mathbb H^{\mc M}$ defined with respect to $\mc M$ into a symmetric Hamiltonian $\mathbb H^{\mc N}$ defined with respect to $\mc N$ \cite{Lootens:2022avn}. Within this formalism, \emph{symmetry}
operators correspond to objects in the fusion category $\mc D^\star_\mc M := \Fun_\mc D(\mc M,\mc M)$, which is referred to as the \emph{Morita dual} of $\mc D$ with respect to $\mc M$ \cite{ETINGOF2011176,10.2140/ant.2013.7.1365,MUGER200381}.

Let us now suppose that the chain is \emph{finite} in length $L+1$. Simply closing the chain by identifying degrees of freedom at sites $1$ and $L+1$ implements \emph{periodic} boundary conditions. More generally, closed boundary conditions can also be twisted by inserting a \emph{symmetry defect} labeled by a simple object in $\mc D^\star_\mc M$, which in contrast to a symmetry operator that extends over the whole space, is localized at a given site and extends in the time direction \cite{Buican:2017rxc,Aasen_2016,Aasen:2020jwb,Lootens:2022avn,Lin:2022dhv}. Concretely, given a boundary condition labeled by $A \in \mc D^\star_\mc M$, the microscopic Hilbert space is given by:
\begin{equation}
    \label{eq:closedHilb}
    \mathbb C \bigg[ \!\! \chain{\, i_{L+\frac{1}{2}}}{i_\frac{1}{2}}{i_\frac{3}{2}
    }{M_L}{M_{L+1}}{M_1}{M_2}{Y_{L+\frac{1}{2}}}{A}{Y_\frac{3}{2}} \!\! \bigg] \, .
\end{equation}
Accordingly, the previous duality operators are promoted to operators $\fr T^{A,B,X,X',k,k'}_{\mc M | \mc N}$ of the form
\begin{equation}
    \label{eq:tube}
    \sum_{\substack{\{Y \in \mc I_\mc D\} \\ \{M \in \mc I_\mc M\} \\ \{N \in \mc I_\mc N\}}}
    \tube{k}{k'}{N_L}{N_{L+1}}{N_1}{N_2}{M_L}{M_{L+1}}{M_1}{M_2}{X}{X'}{X}{B}{A}{Y_{L+\frac{1}{2}}}{Y_\frac{3}{2}} \, ,
\end{equation}
where $A \in \mc I_{\mc D^\star_\mc M}$, $B \in \mc I_{\mc D^\star_\mc N}$ and $X,X' \in \mc I_{\Fun_\mc D(\mc M,\mc N)}$.
In addition to tensors of the form \eqref{eq:omF}, we have introduced tensors whose non-vanishing components evaluate to entries of isomorphisms implementing the composition $X_1 \circ X_2 \to X_3$ of module functors in $\Fun_\mc D(\mc N,\mc O)$ and $\Fun_{\mc D}(\mc M,\mc N)$, respectively. These read
\begin{equation}
    \label{eq:PEPSComp}
    \PEPS{mod}{j}{i}{k}{l}{O}{M}{N}{X_1}{X_2}{X_3}{3} :=
    \big( \F{\fr F}^{X_1 X_2 M}_{O}\big)^{N,lj}_{X_3,ik} \, ,
\end{equation}
where indices $i,j,k,l$ label basis vectors in the hom-spaces $\Hom_{\Fun_\mc D(\mc M,\mc O)}(\Func{X_1}(\Func{X_2}(-)),\Func{X_3}(-)) =: \mc V^{X_3}_{X_1 X_2}$, $\mc V^O_{X_1 N}$, $\mc V^N_{X_2 M}$ and $\mc V^{O}_{X_3 M}$, respectively. It follows that indices $k$ and $k'$ in eq.~\eqref{eq:tube} label basis vectors in  $\mc V^{BX}_{X'} := \Hom_{\Fun_\mc D(\mc M,\mc N)}(\Func{X'}(-),\Func{B}(\Func{X}(-)))$ and $\mc V_{XA}^{X'} := \Hom_{\Fun_\mc D(\mc M,\mc N)}(\Func{X}(\Func{A}(-)),\Func{X'}(-))$, respectively. 

The various pentagon axioms satisfied by the isomorphisms entering the definition of the duality operator \eqref{eq:tube} then give rise to commutation relations of the form
\begin{equation}
    \fr T^{A,B,X,X',k,k'}_{\mc M | \mc N} \circ \mathbb H^{\mc M,A}
    =
    \mathbb H^{\mc N,B} \circ \fr T^{A,B,X,X',k,k'}_{\mc M | \mc N} \, 
\end{equation}
describing the transmutation of symmetric Hamiltonian $\mathbb H^{\mc M,A}$ with boundary condition $A$ into a dual symmetric Hamiltonian $\mathbb H^{\mc N,B}$ with boundary condition $B$ \cite{Lootens:2022avn}. Operators of the form $\fr T_{\mc M|\mc N}$, $\fr T_{\mc N|\mc M}$, $\fr T_{\mc M| \mc M}$ and $\fr T_{\mc N|\mc N}$ span an algebra---referred to as the (enlarged) tube algebra \cite{neshveyev2015few}---with multiplication rule governed by the composition of the corresponding $\mc D$-module functors \cite{Lootens:2022avn}. In particular, the irreducible representations of this algebra provide the superselection sectors $Z$ of the Hamiltonian, which are well known to be in one-to-one correspondence with simple objects in the monoidal center $\mc Z(\mc D^\star_\mc M) \cong \mc Z(\mc D^\star_\mc N) \cong \mc Z(\mc D)$ \cite{ETINGOF2011176,10.2140/ant.2013.7.1365}. Upon dualizing, the sector $Z$ is mapped to a dual sector $\tilde Z$. As an object in $\mc D^\star_\mc M$ (resp. $\mc D^\star_\mc N$), the sector $Z$ (resp. $\tilde Z$) decomposes into simple objects $A_i$ (resp. $B_j$). In our context, the object $A$ (resp. $B$) labels a symmetry twisted boundary condition, and the index $i$ (resp. $j$) labels a component of the corresponding charge.

%% file: _Linear.tex
\medskip \noindent
\textbf{Linear depth circuit:}
We now turn to writing the duality operator of eq.~\eqref{eq:tube} as a quantum circuit. First we introduce two types of gates. On the one hand, we define a \emph{duality gate} as a unitary transformation
\begin{align}
    \gate{0} \; \equiv \; \sum_X \gate{1} \; \equiv \sum_{\substack{X, \\ M_1,M_2 \\ N_1,N_2}} \MPOgate
\end{align}
between the vector spaces $\bigoplus_{X,M_1,M_2,N_2} \mc V_{M_1Y}^{M_2} \otimes \mc V_{X M_2}^{N_2}$ and $\bigoplus_{X,M_1,N_2,N_2} \mc V_{X M_1}^{N_1} \otimes \mc V_{N_1 Y}^{N_2}$. Building a `staircase' quantum circuit out of this gate yields the duality operator labeled by $X$ in the infinite chain case. For the special case $\mc M = \mc N$, we refer to this gate as a \emph{symmetry gate} instead. On the other hand, we define a \emph{composition gate} as a unitary transformation with components
\begin{equation}
    \raisebox{6pt}{\gate{6}} \hspace{-10pt} \equiv 
    \raisebox{6pt}{\gate{2}} 
    \equiv 
    \sum_{\substack{M,N,O}} \, \PEPSgate
\end{equation}
between the vector spaces $\bigoplus_{X',X'',M,N,O} \mc V_{X'M}^{N} \otimes \mc V_{X''N}^{O}$ and $\bigoplus_{X,X',X'',M,X,O} \mc V_{X M}^{O} \otimes \mc V_{X'' X'}^{X}$. Using these two gates, we rewrite the duality operator of eq.~\eqref{eq:tube} as the linear depth quantum circuit depicted in fig.~\ref{fig:linear} as follows:
\begin{figure}
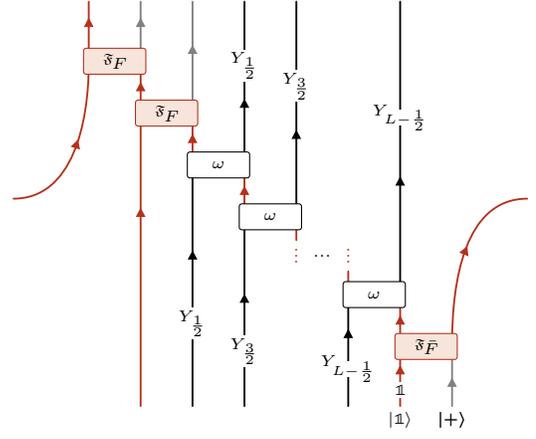

    \circuitClosed
    \caption{Implementation of a general duality with symmetry-twisted closed boundary conditions as a linear depth circuit. Given a state $\ket{\psi}$ with a symmetry twisted boundary condition, we add ancillary states $\ket{\mathbb 1}$ and $\ket{+}$ as defined in the text, before sequentially acting with the duality gate on the physical degrees of freedom $\{Y_{\msf i-1/2}\}_{\msf i=1}^L$. We then act with a composition gate on the symmetry defect strand and the duality strand to its right, and compose the resulting duality strand with an opposite duality strand to obtain a symmetry twist.}
    \label{fig:linear}
\end{figure}
First, we act with the inverse composition gate on an ancillary state $\ket{\mathbb 1} \otimes \ket{+}$, where $\ket{\mathbb 1} \equiv \ket{M_L \mathbb 1 M_L,1}$ and $\ket{+}$ is defined such that\footnote{Notice that the ancillary state $\ket{\mathbb 1}$ shares degrees of freedom with its neighboring sites, reflecting that the Hilbert spaces under consideration are generally not tensor product spaces. This implies that the state $|\mathbb 1\ra$ cannot simply be added via a tensor product. Nevertheless, it can always be created from some reference state via a controlled unitary. Henceforth, we implicitly assume this step whenever we insert such an ancillary state.}
\begin{equation}
    \raisebox{-7pt}{\gate{4}} \equiv 
    \sum_{X \in \mc I_{\Fun_\mc D(\mc M,\mc N)}} \frac{d_X}{\text{FPdim}(\mc D)} \raisebox{-7pt}{\gate{5}} \hspace{-5pt} ,
\end{equation}
where $\bar X \in \mc I_{\Fun_\mc D(\mc N,\mc M)}$ labels the unique `opposite' duality strand such that its composition with $X$ contains the identity in $\mc D_\mc M^\star$. Furthermore, $d_X$ is the dimension of $X$ as a simple object in the $\mc D_\mc M^\star$-module category $\Fun_\mc D(\mc M,\mc N)$ and $\text{FPdim}(\mc D)$ is the Frobenius-Perron dimension of $\mc D$ \cite{etingof2016tensor}, which ensures that $| + \ra$ is normalized.

We can now implement the action of the duality on the physical degrees of freedom $\{Y_{\msf i-1/2}\}_{\msf i=1}^L$ by acting with a sequence of composition and duality gates as shown in fig.~\ref{fig:linear}. Explicitly, the action of this circuit $U$ on a state $\ket{\psi}$ in the Hilbert space \eqref{eq:closedHilb} can be written as
\begin{align}
    \label{eq:explicit_linear}
    \!\!\!U \big(\ket{\psi} \otimes \ket{\mathbb 1} \otimes \ket{+} \big) 
    =  
    \!\! \sum_{\substack{k,k',B,A\\ X,X'}} \! 
    \big(&\tilde{\fr T}_{\mc M|\mc N}^{A,B,X,X',k,k'}
    \ket{\psi} \big)
    \\[-1.2em] \nn
    &\!\!\! \otimes \ket{\bar{X}X'B,k} \otimes 
    \ket{AXX',k'}
\end{align}
where the tubes $\tilde{\fr T}$ are related to $\fr T$ by a linear transformation between $\mc V_{B X}^{X'}$ and $\mc V^{X'\bar X}_B$. These also form a basis of the tube algebra, with the additional property of being unitary as matrices on the Hilbert space. The circuit produces additional degrees of freedom $\ket{\bar{X}X'B,k}$ and $\ket{AXX',k'}$; to interpret them, we define a unitary transformation $\Omega$ via
\begin{equation*}
    \sqrt{\frac{d_B}{d_{\tilde Z}}} e^{Z,A_i,\tilde Z,B_j}_{\mc M | \mc N} = \!\!\!\! \sum_{X,X',k,k'} \!\!\!\!\!\! \Omega^{Z,A_i,\tilde Z,B_j}_{X,X',k,k'} \, \tilde{\fr T}_{\mc M|\mc N}^{A,B,X,X',k,k'},
\end{equation*}
where $e^{Z,A_i,\tilde Z,B_j}_{\mc M | \mc N}$ are the matrix units that diagonalize the multiplication rule of the tube algebra \cite{Lootens:2022avn}. The matrix units are labeled by a component $A_i$ in the topological sector $Z \in \mc Z(\mc D^\star_\mc M)$ and a dual component $B_j$ in the corresponding dual sector $\tilde Z \in \mc Z(\mc D^\star_\mc N)$. Acting with $\Omega$ on $\ket{\bar{X}X'B,k}$ and $\ket{AXX',k'}$, we find
\begin{align}
    &\Omega \, U \big(\ket{\psi} \otimes \ket{\mathbb 1} \otimes \ket{+} \big) 
    \\ \nn 
    & \q =  
    \sum_{\substack{Z \\ A_i,B_j}} \sqrt{\frac{d_B}{d_{\tilde Z}}} \big(e^{Z,A_i,\tilde Z,B_j}_{\mc M | \mc N}\ket{\psi}\big) 
    \otimes \ket{Z,A_i} \otimes \ket{\tilde Z,B_j}.
    \label{eq:linear}
\end{align}
Concretely, this signifies that measuring the ancillary states $\ket{Z,A_i}$ and $\ket{\tilde Z,B_j}$ amounts to implementing the duality operator proportional to $e^{Z,A_i,\tilde Z,B_j}_{\mc M | \mc N}$, which maps the component $A_i$ of the sector $Z$ to the component $B_j$ of the dual sector $\tilde Z$. We note that if all sectors $Z$ and $\tilde Z$ happen to be simple as objects of $\mc D^\star_\mc M$ and $\mc D^\star_\mc N$, the ancillary states can be disentangled in linear depth by acting with the Hermitian conjugate of $\Omega \, U$ where we now choose $\mc M = \mc N$ instead. 

%% file: _Constant.tex
\medskip \noindent
\textbf{Constant depth circuit for nilpotent dualities:} 
It was recently appreciated that certain dualities can be implemented in constant depth, provided that the unitary quantum circuit is supplemented with local operations and classical communication (LOCC) \cite{PhysRevLett.127.220503,PRXQuantum.4.020339,Bravyi:2022zcw}. In (1+1)d, we show that this is possible whenever the dual symmetry fusion category $\mc D _\mc N^\star$ is \emph{nilpotent}, through a procedure similar to the one of \cite{Dauphinais_2017,galindo2018acyclic}. A \emph{graded} fusion category is a fusion category of the form $\mc C = \bigoplus_{g\in G} \mc C_g$ with $G$ a finite group, such that the product rule $\otimes: \mc C_g \times \mc C_h \rightarrow \mc C_{gh}$ respects the group multiplication. We can define a sequence of fusion categories $\mc C =: \mc C^{(0)}, \mc C^{(1)},\ldots , \mc C^{(n)}$ and a sequence of finite groups $G^{(0)}, G^{(1)}, \ldots, G^{(n)}$ such that $\mc C^{(i)}$ is (faithfully) graded by $G^{(i)}$ and the trivial component of $\mc C^{(i-1)}$ is $\mc C^{(i)}$, for all $i = 1,\ldots,n$. If there exists a sequence such that $\mc C^{(n)} = \Vect$---the unique fusion category with a single simple object---then $\mc C$ is said to be nilpotent and the length $n$ of the shortest such sequence is its nilpotency class \cite{GELAKI20081053,ETINGOF2011176}. In particular, it follows from the definition that $\mc C^{(n-1)}$ must be of the form $\Vect^\omega_{G^{(n-1)}}$, capturing an invertible $G^{(n-1)}$-symmetry with 't Hooft anomaly $[\omega] \in H^3(G^{(n-1)},{\rm U}(1))$. Given $\mc C^{(i)}$ and a simple object $B$, we denote its homogeneous component of degree $g \in G^{(i)}$ by $B^{(i)}_g$.

\begin{figure}
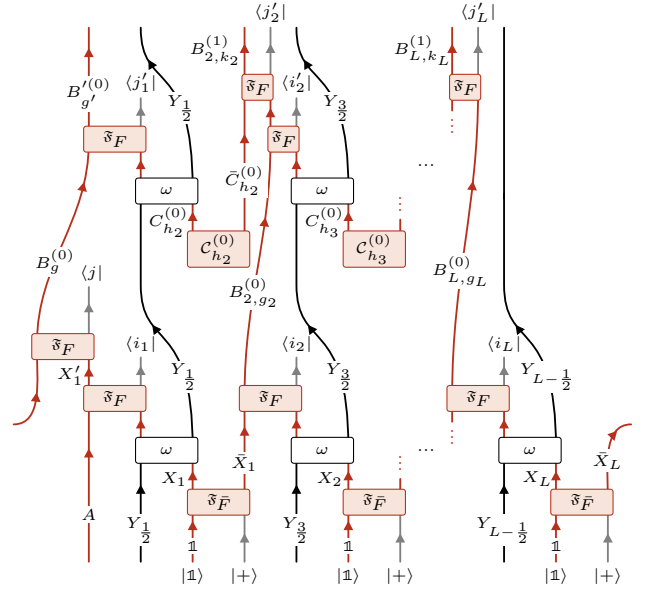

    \centering
    \circuitConstant
    \caption{Implementation of a nilpotent duality as a constant depth circuit. Given an initial state with some symmetry twisted boundary condition, we begin by adding ancillary states $\ket{\mathbb 1}$ and $\ket{+}$ in between every physical site before acting with inverse composition gates, followed by the action of duality gates. Then, the boundary condition is composed with its neighboring duality strand while the corresponding multiplicity degree of freedom $i_1$ is measured. The first stage is concluded by composing all adjacent duality strands and measuring the resulting multiplicity degrees of freedom resulting in symmetry twists in $\mc D^\star_\mc N =: \mc C^{(0)} $. The second stage amounts to inserting new ancillary states to the left of the symmetry twists via \eqref{eq:cupC}, before acting with symmetry gates. The second stage is concluded by composing all adjacent symmetry strands and measuring the resulting multiplicity degrees of freedom, resulting almost exclusively in symmetry twists in $\mc C^{(1)}$. Repeating this process $n-1$ more times results in trivial symmetry twists---apart from the leftmost one---labeled by the single simple object in $\mc C^{(n)} = \Vect$.
    }
    \label{fig:constant}
\end{figure}
Given such a nilpotent duality, the corresponding constant depth circuit is sketched in fig.~\ref{fig:constant}. It consists of $n+1$ stages. For the first stage, we start by adding the previously defined ancillary state $\ket{\mathbb 1} \otimes \ket{+}$ in between every site before acting with a sequence of composition and duality gates, concluded by the measurements of multiplicity degrees of freedom. The result is the implementation in constant depth of a modified version of the duality operator where symmetry twists $B^{(0)}_{\msf i,g_\msf i} \in \mc C^{(0)} := \mc D_\mc N^\star$ are situated in between every two sites. The goal is then to compose all these symmetry twists into a single one so as to recover the original Hilbert space. Since $\mc C^{(0)}$ is nilpotent this can be done in $n$ stages via constant depth circuits and measurements.  
Concretely, we define ancillary states
\begin{equation}
    \label{eq:cupC}
    \raisebox{8pt}{\gate{7}} \equiv \sum_{C^{(i)}_h \in \mc I_{\mc C^{(i)}_h}} \frac{d_{C^{(i)}_h}}{\text{FPdim}\big(\mc C^{(i)}_h\big)} \gate{8},
\end{equation}
for every $i = 1, \ldots,n$ and $h \in G^{(i)}$. These are inserted in between physical sites alongside symmetry twists $B^{(0)}_{\msf i,g_\msf i}$, $\msf i = 2,3,\ldots,L$ such that $h_\msf i = g_L g_{L-1} \ldots g_{\msf i}$, where the group elements $g_\msf i$ were obtained from the measurements in the previous stage (see fig.~\ref{fig:constant}). We subsequently act with symmetry and composition gates, before measuring the resulting multiplicity degrees of freedom. Due to the choice of variables $h_\msf i$, $\msf i=2,\ldots,L$, the resulting symmetry strands are now labeled by objects $B^{(1)}_{\msf i,g_\msf i} \in \mc C^{(1)}$, aside from the first symmetry twist which remains in $\mc C^{(0)}$. Repeating this process $n-1$ more times eventually yields a single non-trivial symmetry twist, as desired.

Note that the operator implemented by the circuit of fig.~\ref{fig:constant} is equivalent to a linear combination of operators of the form $\fr T_{\mc M | \mc N}$, which can be obtained by invoking associativity of composition of module functors \cite{Lootens:2022avn}. Since operators $\fr T_{\mc M | \mc N}$ are diagonal in the topological sector $Z$ and its dual $\tilde Z$, this constant depth circuit maps components $A_i$ of $Z$ into a superposition of components $B_i$ of $\tilde Z$, with amplitudes determined by the measurement outcomes and the initial state.

%% file: _Discussion.tex
\medskip \noindent
\textbf{Discussion}
We have shown that general dualities for (1+1)d quantum lattice models can be implemented with linear depth unitary quantum circuits when supplemented with ancillary states. In the case that all sectors are one-dimensional, the ancillary states can be disentangled by acting with a self-duality in linear depth, such that the duality acts as a unitary. For nilpotent dualities, we demonstrated that the duality transformation can be implemented in constant depth by including measurements. If the symmetries of the dual model are given by a finite abelian group, circuits implementing such dualities have been constructed in ref.~\cite{PRXQuantum.4.020339,PRXQuantum.4.020315,PhysRevB.106.085122}, and one can check that our general construction reproduces these special cases. Furthermore, we note that the condition of nilpotency for constant depth is too strong; it is sufficient that the bulk symmetry twists created by the measurements are nilpotent, which is less stringent than requiring the full symmetry category to be nilpotent. We leave a full characterization of this more general case to future work.

The advantage of the categorical formulation of dualities employed here is that it can readily be generalized to (2+1)d \cite{Delcamp:2021szr,Delcamp:2023kew}, although a general treatment of the interplay with superselection sectors has not yet been worked out. It has been argued that (2+1)d topologically ordered phases based on solvable fusion categories can be prepared with a constant depth circuit and LOCC through sequential gauging \cite{williamson2017symmetryenriched,PRXQuantum.4.020339}. Recently, these circuits have been used to provide the first experimental realization of the ground state of a topological phase exhibiting non-abelian anyons \cite{Iqbal:2023wvm}, and we expect the higher dimensional realization of our result to contribute to the ongoing effort of realizing exotic phases of matter in quantum devices.